\documentclass[conference,compsocconf]{IEEEtran}

\IEEEoverridecommandlockouts
\usepackage{placeins}

\usepackage{cite}
\usepackage{amsmath,amssymb,amsfonts}
\usepackage{algorithmic}
\usepackage{graphicx}
\usepackage{textcomp}
\usepackage{xcolor}
\usepackage{enumitem}
\usepackage{colortbl}

\usepackage{hyperref}
\usepackage{subcaption}
\def\BibTeX{{\rm B\kern-.05em{\sc i\kern-.025em b}\kern-.08em
    T\kern-.1667em\lower.7ex\hbox{E}\kern-.125emX}}

\usepackage{etoolbox}
\apptocmd{\thebibliography}{\clearpage}{}{}

\renewcommand{\baselinestretch}{0.948}

\begin{document}

\title{MORPH: Multi-Environment Orchestrated Reinforcement Learning for PRB Handling in O-RAN \thanks{This material is based upon work supported by the National Science Foundation under Grant Numbers  CNS-2202972, CNS- 2318726, and CNS-2232048 and Clemson R-Initiative Grant.}
}
% \author{Anonymous}

\author{
\IEEEauthorblockN{ Alireza Ebrahimi Dorcheh, Tolunay Seyfi, Ryan Barker, Fatemeh Afghah}
\IEEEauthorblockA{
\textit{Holcombe Department of Electrical and Computer Engineering} 
\textit{Clemson University},
Clemson, South Carolina, USA \\
\{alireze,tseyfi,rcbarke,fafghah\}@clemson.edu}
%\and

% \IEEEauthorblockN{2\textsuperscript{nd} Tolunay Seyfi}
% \IEEEauthorblockA{
% \textit{Holcombe Department of Electrical} \\
% \textit{and Computer Engineering} \\
% \textit{Clemson University} \\
% Clemson, South Carolina, USA \\
% tseyfi@clemson.edu}
% \and

% \IEEEauthorblockN{3\textsuperscript{rd} Ryan Barker}
% \IEEEauthorblockA{
% \textit{Holcombe Department of Electrical} \\
% \textit{and Computer Engineering} \\
% \textit{Clemson University} \\
% Clemson, South Carolina, USA \\
% rcbarke@clemson.edu}
% \and

% \IEEEauthorblockN{4\textsuperscript{th} Fatemeh Afghah}
% \IEEEauthorblockA{
% \textit{Holcombe Department of Electrical} \\
% \textit{and Computer Engineering} \\
% \textit{Clemson University} \\
% Clemson, South Carolina, USA \\
% fafghah@clemson.edu}
}

\maketitle

\begin{abstract}

Reinforcement-learning (RL) solutions for dynamic spectrum access and radio resource management in Open Radio Access Networks (O-RAN) depend critically on the fidelity of the throughput signal used for training. Analytical or physical-layer (PHY)-only simulators scale well but often miss protocol-stack effects such as signaling overhead and retransmissions, whereas exhaustive throughput profiling on a standards-compliant 5G stack is slow and can be unstable under software execution constraints.
This paper presents MORPH, a measurement-grounded multi-environment RL pipeline 
%for slice-aware spectrum allocation at physical resource block (PRB) granularity
{for slice-aware PRB-level spectrum allocation (spectrum sharing and slice isolation within a single gNB)}
built on OpenAirInterface (OAI) 5G-NR RF-simulator mode. MORPH leverages three complementary throughput sources: (i) application-layer throughput measured via \texttt{iPerf} on the OAI stack under controlled AWGN pathloss settings, (ii) empirical MCS-selection distributions conditioned on path loss, enabling a distribution-aware theoretical throughput estimator that reflects standards-compliant link adaptation, and (iii) scalable throughput estimates from a 3GPP-parameterized PHY-fidelity OFDM simulator.
Using these components, we train and compare agents that differ only in the origin of their throughput feedback: an OAI-grounded practical agent, a simulator-driven agent, and MORPH, which fuses real and synthetic throughput signals for policy optimization. %Evaluation on the OAI execution harness across heterogeneous slicing scenarios shows that MORPH yields more robust slice-wise performance and improved SLA compliance than single-source training, highlighting the trade-off between testbed fidelity and data-collection scalability.
{Evaluation on the OAI execution harness across heterogeneous slicing scenarios shows that MORPH yields more robust slice-wise performance and improved SLA compliance than single-source training, providing a practical foundation for PRB-level spectrum sharing and slice isolation within a single-cell stack and a stepping stone toward multi-cell spectrum coordination and interference management.}

\end{abstract}

\begin{IEEEkeywords}
 Open RAN, reinforcement learning, network slicing, OpenAirInterface, radio resource management, AI-native RAN.
\end{IEEEkeywords}

\section{Introduction}
The Open Radio Access Network (O-RAN) paradigm represents a transformative shift in the design of next-generation wireless systems, moving away from rigid, vendor-locked architectures toward openness, disaggregation, and intelligence. By enabling a modular and standards-compliant ecosystem, O-RAN promotes interoperability and innovation across the RAN stack~\cite{10024837}.

A cornerstone of this vision is the RAN Intelligent Controller (RIC), which enables closed-loop control via xApps and rApps for functions such as Physical Resource Block (PRB) allocation, interference management, and mobility optimization~\cite{bonati2020intelligence}. These capabilities are critical for supporting the heterogeneous Quality of Service (QoS) requirements of enhanced Mobile Broadband (eMBB), Ultra-Reliable Low-Latency Communication (URLLC), and massive Machine-Type Communication (mMTC)~\cite{lopez2020towards,Lotfi25}.

Despite this progress, achieving real-time adaptive resource allocation in O-RAN remains challenging. Reinforcement Learning (RL) has emerged as a promising approach for dynamic PRB allocation and network slicing due to its ability to learn optimal policies under uncertainty. However, deploying RL-based control in O-RAN faces three key challenges: (i) maintaining alignment between application-layer QoS objectives and low-level resource allocation decisions in disaggregated architectures~\cite{Kouchaki2022, Ebrahimi25}, particularly when control is mediated through RIC-based xApps; (ii) acquiring large-scale, structured training data from live networks, which is hindered by limited observability and telemetry noise~\cite{Mollahasani2021}; and (iii) mitigating the simulation-to-reality gap, as policies trained solely in simulators often fail in deployment due to unmodeled protocol overhead, retransmissions, and CPU-induced processing bottlenecks~\cite{zhang2022sim2real,wagenmaker2024overcoming,dorcheh2025_dora}.
% \fa{cite DORA, BASIR WCNC and Fatemeh's ICC's paper. it's not double-blind anymore.}

%% Basir Start
 To address these challenges, we propose MORPH (Multi-Environment Orchestrated Reinforcement Learning for PRB Handling), a measurement-grounded training pipeline for slice-aware PRB allocation that focuses on the \emph{fidelity and coverage of the throughput learning signal}. MORPH has three components. First, we collect application-layer throughput measurements from an OAI 5G stack via \texttt{iPerf}, capturing stack-level effects such as signaling overhead, retransmissions, and software processing artifacts that are typically absent from custom simulators. Second, we develop a 3GPP-parameterized PHY-fidelity OFDM simulator to enable scalable exploration of PRB allocations and channel-quality regimes that would be prohibitively slow to sweep exhaustively in OAI. Third, we construct a hybrid throughput oracle that combines testbed-derived and simulator-derived throughput estimates, enabling off-testbed policy optimization while remaining grounded in OAI behavior. For benchmarking, we consider two single-source baselines: training with OAI-derived throughput using proportional PRB scaling, and training using the OFDM simulator with empirical MCS-weighting. 

%% Basir End

We validate MORPH across three representative deployment scenarios—smart factory automation, large-scale entertainment venues, and smart city traffic control—spanning diverse traffic patterns and QoS objectives. Experimental results show that MORPH significantly improves robustness over purely simulation-trained and purely OAI-trained agents, effectively narrowing the simulation-to-reality gap while preserving standards compliance. 
%% Basir Start
To the best of our knowledge, this is among the first OAI-based studies that (i) does a throughput analysis on OAI RF simulator, (ii) extracts empirical MCS selection distributions versus pathloss from a standards-compliant 5G stack using AWGN channel modeling, (iii) uses these distributions to build a distribution-aware throughput estimator validated against application-layer measurements, and (iv) leverages these multi-source throughput models as an offline training signal for slice-aware PRB allocation.
% \fa{Great statement, Basir!}

\noindent\textbf{Scope and comparison philosophy.} Our objective is not to propose a new RL architecture for PRB allocation, but to quantify and mitigate the throughput-model mismatch that arises when learning signals are obtained from either (i) protocol-stack measurements with limited coverage, or (ii) simulators with limited fidelity. Accordingly, our baselines are intentionally constructed to differ \emph{only} in the source of the throughput signal (OAI-only vs. simulator-only), allowing us to attribute performance differences directly to training-data realism and coverage rather than to unrelated algorithmic changes.
%% Basir End

\section{Related Work}

The integration of RL into Open RAN (O-RAN) architectures has emerged as a promising approach for intelligent radio resource management. The disaggregated and programmable nature of O-RAN enables AI-driven xApps and rApps within the RIC, supporting closed-loop control over key radio functions. Among these, PRB allocation is central, as it directly impacts throughput, latency, interference management, and service-level agreement (SLA) compliance across heterogeneous QoS slices~\cite{10329947,10625184}.

\subsection{RL-Based PRB Allocation in O-RAN}

Deep reinforcement learning (DRL) has been widely explored for PRB allocation due to its ability to adapt to time-varying channel conditions and heterogeneous traffic demands. Simulation- and emulation-based frameworks such as PandORA~\cite{tsampazi2025pandora} and ColO-RAN~\cite{polese2022colo} evaluate DRL-driven xApps under varying reward structures, action granularities, and traffic models, demonstrating the potential of learning-based spectrum management. Related orchestration frameworks, including Oranus~\cite{adamuz2024oranus}, which leverages stochastic network calculus for latency-aware control, and SEM-O-RAN~\cite{puligheddu2023sem}, which introduces semantic slicing for edge-assisted systems, further advance algorithmic optimization within the O-RAN ecosystem. However, these approaches predominantly emphasize the learning or optimization algorithm itself and rely on abstracted environments that omit key real-world effects, including protocol signaling overhead, retransmission behavior, scheduler implementation details, and control–data plane interactions. As a result, a gap remains between simulated rewards and the physical constraints imposed by the O-RAN protocol stack—a disconnect that motivates a shift toward higher-fidelity learning environments.

To improve realism, several works adopt OAI-based testbeds. ORANSlice~\cite{cheng2024oranslice} introduces a standards-compliant slicing framework built on OpenAirInterface (OAI), enabling multi-slice operation through enhanced control-plane mechanisms. While ORANSlice supports realistic slice-level experimentation with both hardware and software UEs, its control logic is static and rule-based, without learning or adaptive optimization. Moreover, its evaluation considers limited UE configurations and does not address policy generalization under dynamic traffic conditions.

Complementary system-level approaches emphasize orchestration rather than PRB-level learning. OrchestRAN~\cite{doro2024orchestrran} studies AI model placement across distributed O-RAN nodes under latency and compute constraints, while AdaSlicing~\cite{zhao2025adaslicing} combines Bayesian optimization and ADMM-based coordination to improve inter-slice isolation and spectrum efficiency. Although these systems demonstrate sophisticated orchestration strategies, their evaluations remain largely simulation-driven and do not explicitly consider link-layer dynamics or end-to-end throughput behavior.

Overall, prior RL-based O-RAN studies primarily operate in synthetic or partially emulated environments, limiting their ability to capture protocol-level inefficiencies and processing bottlenecks that dominate real-world performance.

\subsection{Throughput Regimes and Control Granularity in O-RAN Systems}

Several recent OAI-based systems evaluate slicing and control under constrained radio configurations. In~\cite{McManus24}, experiments use a 10~MHz carrier (50~PRBs) with up to three UEs, resulting in aggregate downlink throughput consistently below 30~Mbit/s. This narrowband regime limits the PRB allocation space, such that slicing decisions primarily redistribute a fixed throughput budget rather than explore fine-grained control at scale. A related line of work in~\cite{Moro23} introduces an xApp for SLA enforcement under guaranteed bit-rate constraints. Although a wider carrier is configured, the effective radio budget is explicitly capped, and throughput serves mainly as a compliance metric rather than a learning signal. Similarly, HexRAN~\cite{Kak25} evaluates architectural scalability under traffic-limited conditions, where throughput closely tracks offered load, demonstrating system correctness but not PRB-aware optimization. TC-RAN~\cite{Irazabal24} prioritizes latency and flow isolation, deliberately trading off throughput stability and leaving radio resources idle to meet delay targets.

Even systems that configure substantially wider carriers exhibit similar limitations. x5G evaluates slicing under \cite{x5G} 100~MHz (273~PRBs) deployments at identical numerology; however, observed downlink throughput remains largely around 50 Mbps. While PRB allocations scale correctly with bandwidth, throughput does not, indicating a traffic- or system-limited regime rather than a PRB-limited one. Consequently, slicing decisions effectively redistribute a bounded throughput budget, and wideband PRB-level control is not exercised.

Across these systems, throughput is predominantly treated as an evaluation outcome or constrained variable rather than as a learnable, PRB-dependent quantity. As a result, the behavior of control policies in wideband, high-dimensional PRB allocation spaces remains largely unexplored. 
% \fa{Great!}

\subsection{Bridging the Simulation-to-Reality Gap in O-RAN}

Despite strong simulation performance, deploying DRL policies in real O-RAN systems remains challenging due to the simulation-to-reality gap. Custom simulators often abstract key effects such as Hybrid Automatic Repeat reQuest (HARQ) retransmissions, control signaling overhead, and processing delays, leading to poor generalization in deployment~\cite{Kouchaki2022,zhang2022sim2real,wagenmaker2024overcoming}. Conversely, training directly on testbeds is hindered by limited observability, slow convergence, and operational instability.

Several works attempt to mitigate this gap. SafeSlice~\cite{nagib2025safeslice} introduces risk-sensitive rewards and safety layers to limit SLA violations under dynamic traffic but remains confined to virtualized environments. OnSlicing~\cite{liu2021onslicing} applies online DRL within an OAI-based testbed, enabling direct adaptation at the cost of slow convergence and high exploration overhead. Regression-assisted and domain-randomized approaches~\cite{sulaiman2023generalizable} improve robustness to unseen traffic but rely on synthetic training data. The REAL framework~\cite{barker2025real} deploys a closed-loop RL system on a standards-compliant RIC using srsRAN, yet follows a fully online paradigm without leveraging offline data or PHY-compliant simulation.

These approaches highlight a fundamental trade-off: simulation-driven methods lack fidelity, while testbed-driven methods suffer from limited coverage and efficiency.

\subsection{Learning-based SLA-aware Control with MORPH}
\label{subsec:novelty}

MORPH addresses this gap through a hybrid training paradigm for slice-aware PRB allocation that jointly leverages empirical OAI measurements and synthetic feedback from a 3GPP-compliant physical-layer simulator. Unlike prior systems operating in narrowband or throughput-capped regimes, MORPH operates at full wideband resolution (106~PRBs, 30~kHz subcarrier spacing), reflecting practical mid-band NR deployments and significantly increasing control complexity.

A key distinction of MORPH is the use of application-layer throughput as a first-class learning signal. Throughput is measured directly from the OAI stack, capturing protocol overheads, HARQ behavior, and scheduler dynamics absent from synthetic models. To overcome limited observability on live testbeds, these measurements are complemented by a PHY-faithful simulator that enables systematic exploration of PRB allocations and channel conditions impractical to encounter online.

MORPH further integrates throughput learning with explicit SLA modeling across heterogeneous service classes, including eMBB throughput guarantees and URLLC latency constraints. Rather than enforcing static guarantees or optimizing a single metric, MORPH learns to dynamically balance throughput efficiency and SLA satisfaction as network conditions evolve.

\subsection{Contributions and Distinguishing Factors}

The main contributions of this work are summarized as follows:
\begin{itemize}[left=0pt]
    \item \textbf{Application-Layer Throughput as a Learning Signal:} We conduct a comprehensive throughput analysis on an OAI-based testbed using downlink \texttt{iPerf}, capturing protocol overheads, retransmissions, and scheduling delays that impact user-perceived performance.
    \item \textbf{3GPP-Compliant Link-Layer Simulation:} We develop a high-fidelity OFDM link-layer simulator modeling PRB allocation, modulation, channel conditions, and link adaptation under 3GPP-compliant parameters.
    \item \textbf{Hybrid Sim--Real Training Framework:} We introduce MORPH, a hybrid RL agent trained using both empirical OAI measurements and interactive simulator queries, and benchmark it against purely real-data and purely simulation-trained agents.
    \item \textbf{Wideband, SLA-Aware PRB Control:} MORPH operates at full wideband resolution and jointly optimizes throughput and SLA satisfaction across heterogeneous slices, moving beyond narrowband or throughput-capped regimes.
    \item \textbf{Scenario-Driven Evaluation:} We validate MORPH across representative deployment scenarios, including smart factories, large-scale entertainment venues, and smart city traffic control.
\end{itemize}

\section{System and Simulation Model}

\subsection{OAI Testbed and Slice-Aware PRB Control}

\begin{table*}[ht]
\centering
\definecolor{headerblue}{RGB}{0, 51, 102}
\definecolor{rowblue}{RGB}{230, 240, 255}
\definecolor{textblue}{RGB}{0, 51, 102}
\caption{Comparison of Training Environments Used in MORPH}
\renewcommand{\arraystretch}{1.3}
\begin{tabular}{|>{\columncolor{rowblue}}p{3.8cm}|p{4.0cm}|p{4.0cm}|p{4.0cm}|}
\rowcolor{headerblue}
\textcolor{white}{\textbf{Feature}} &
\textcolor{white}{\textbf{OAI-Based Emulated Environment}} &
\textcolor{white}{\textbf{PHY-Fidelity Simulator}} &
\textcolor{white}{\textbf{MORPH}} \\
\hline
Execution Mode & Emulated full RAN stack with serialized UEs & Custom C++ PHY-layer simulator & Hybrid \\
\hline
Realism (Protocol Stack) & Full 5G stack (RRC, PDCP, RLC, MAC, PHY), with E2-inspired control interface & PHY layer only (OFDM, MCS) & Captures both high- and low-level effects \\
\hline
Bottleneck Modeling & Yes—protocol signaling, retransmissions, CPU load & No—idealized and deterministic & Improved realism and generalization \\
\hline
Throughput Metric & Application-layer QoE (iPerf) & PHY-layer throughput & Combined for robust reward shaping \\
\hline
State Exploration & Limited to logged scenarios & Fully interactive & Real + synthetic transitions \\
\hline
Scalability & Limited by OAI CPU and serialization & Scalable without real-time constraints & Balanced realism and scalability \\
\hline
Use in MORPH & Realistic slice-aware protocol effects & Wide MCS regime exploration & Generalizable, grounded policies \\
\hline
\end{tabular}
\label{tab:env_comparison}
\end{table*}

\begin{figure}[t]
    \centering
    \includegraphics[width=\columnwidth]{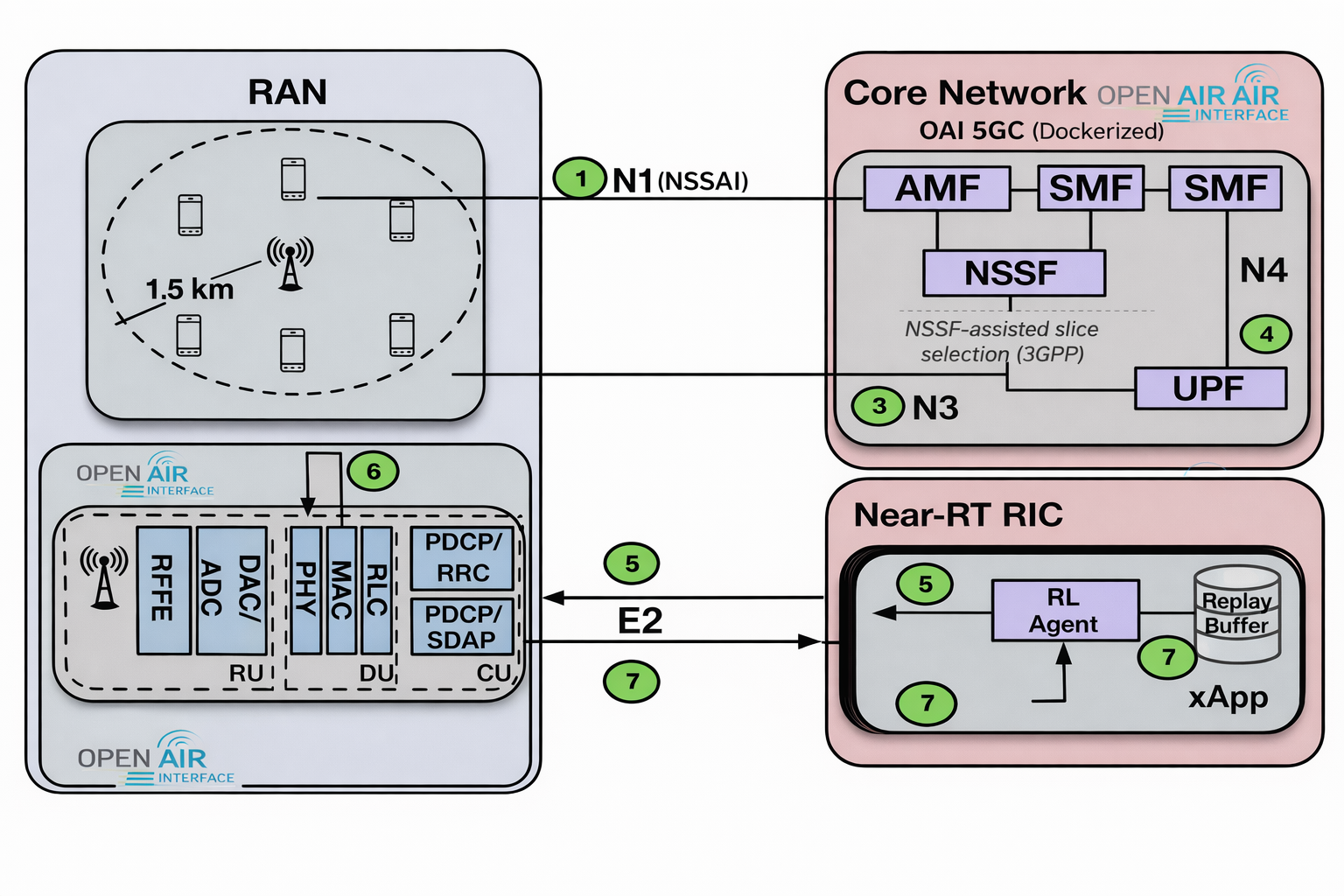}
    \caption{{\small Slice-aware PRB allocation framework across the O-RAN core, RAN, and Near-RT RIC. MORPH deploys the full slicing-capable OAI 5GC via Docker, including the Network Slice Selection Function (NSSF), such that slice selection follows the 3GPP NSSF-assisted model using UE-provided NSSAI. All other UE registration, PDU session establishment, and control/user-plane signaling procedures remain standard-compliant and unchanged relative to minimalist AMF/SMF/UPF deployments.}}
    \label{fig:ran_architecture}
    \vspace{-5pt}
\end{figure}

We first describe the OAI-based environment used for measurement-grounded evaluation in MORPH. As shown in Fig.~\ref{fig:ran_architecture}, policy optimization is performed off-testbed, while OAI is used to (i) collect application-layer throughput measurements and (ii) execute a controlled evaluation harness. Due to software baseband constraints, concurrent multi-UE execution can introduce CPU-induced artifacts that distort throughput measurements. We therefore adopt a \emph{serialized execution} model in which UEs are activated one at a time to obtain stable and repeatable application-layer throughput using \texttt{iPerf}. We emphasize that serialization intentionally removes multi-user contention and interference dynamics from the OAI execution path. These effects are instead captured at the resource-budget level in the simulator-driven components, and we discuss the resulting generalization implications in Section~\ref{sec:limitations}.

During each serialized run, a single UE actively transmits while retaining its full slice association, protocol data unit (PDU) session, and QoS configuration. Although only one UE is active, slice contexts for all users remain instantiated, preserving scheduler behavior and slice-aware policy enforcement. This design enables fair comparison with simulator-based evaluations while maintaining architectural fidelity within the OAI stack.

The serialized execution workflow proceeds as follows.

\textbf{Step 0 (Core bootstrapping: network slice selection assistance information (NSSAI) availability).}
As part of slice availability management, the AMF provisions and maintains slice support information per Tracking Area by updating the NSSF with supported S-NSSAIs via \emph{Nnssf\_NSSAIAvailability} and subscribing to availability change notifications. This establishes slice feasibility prior to UE registration.

\textbf{Step 1 (Slice selection over N1 with NSSF assistance).}
During initial registration, each UE includes a Requested NSSAI in the NAS Registration Request. The AMF retrieves slice subscription information from the unified data management (UDM), including Subscribed S-NSSAIs, default slice indicators, and any simultaneous-registration constraints. When required, the AMF invokes the NSSF (\emph{Nnssf\_NSSelection}) with the Requested NSSAI, subscription data, and the UE’s current Tracking Area. Based on slice availability and policy, the NSSF determines the Allowed NSSAI, optionally providing mappings to subscribed S-NSSAIs and slice instance identifiers. The AMF remains the registration anchor, completing UE registration, returning the Allowed NSSAI to the UE, and conveying slice information to the RAN over N2.

\textbf{Step 2 (Control-plane signaling over N2).}
Slice identity and associated QoS parameters are delivered to the gNodeB via standard N2 procedures (e.g., Initial Context Setup). Under serialized execution, these control-plane contexts persist unchanged, although only the active UE generates MAC-layer traffic.

\textbf{Step 3 (User-plane connectivity over N3).}
GTP-U tunnels between the gNodeB and UPF are established and maintained for all active PDU sessions. Serialization restricts data forwarding to a single UE without altering tunnel mappings, QoS identifiers, or control-plane state.

\textbf{Step 4 (UPF configuration via N4).}
The SMF installs slice-specific enforcement and QoS rules at the UPF using N4 signaling. These rules persist across serialized runs, ensuring consistent slice-level treatment independent of UE execution order.

\textbf{Step 5 (State observation and xApp decision process).}
Serialization primarily affects the observable system state during online evaluation. While parallel execution would expose multi-UE contention through E2 telemetry, the serialized setup collapses the observed state $s[t]$ to that of a single UE, yielding a reduced-complexity operating regime for the Near-RT RIC. To preserve realism, we adopt a hybrid evaluation strategy in which a Python-based simulator generates mobility, traffic demand, and per-timestep pathloss values that are injected into the OAI testbed via TELNET, while application-layer throughput is measured using \texttt{iPerf}. Training remains fully off-testbed using either OAI-derived logs or a simulator-driven environment that captures inter-slice coupling through the shared PRB budget and scenario-dependent reward objectives.

\textbf{Step 6 (PRB enforcement at the MAC scheduler).}
The proportional-fair (PF) MAC scheduler enforces per-UE PRB caps derived from slice association and QoS configuration. Even in the absence of real-time contention, slice priorities and scheduling weights remain active, preserving standards-compliant scheduling behavior.

\textbf{Step 7 (Telemetry feedback and policy execution).}
Although the full O-RAN E2 stack is not instantiated, we employ an E2-inspired control architecture. A Python controller interfaces with OAI through shared memory and internal APIs, enabling low-latency injection of telemetry (e.g., pathloss) and traffic demand profiles while maintaining a clear separation between control and data planes, consistent with prior lightweight E2-based approaches.

\subsection{Synthetic Training Environment for Simulation-Based Agent}

\begin{figure*}[t]
    \centering
    \includegraphics[width=0.7\textwidth]{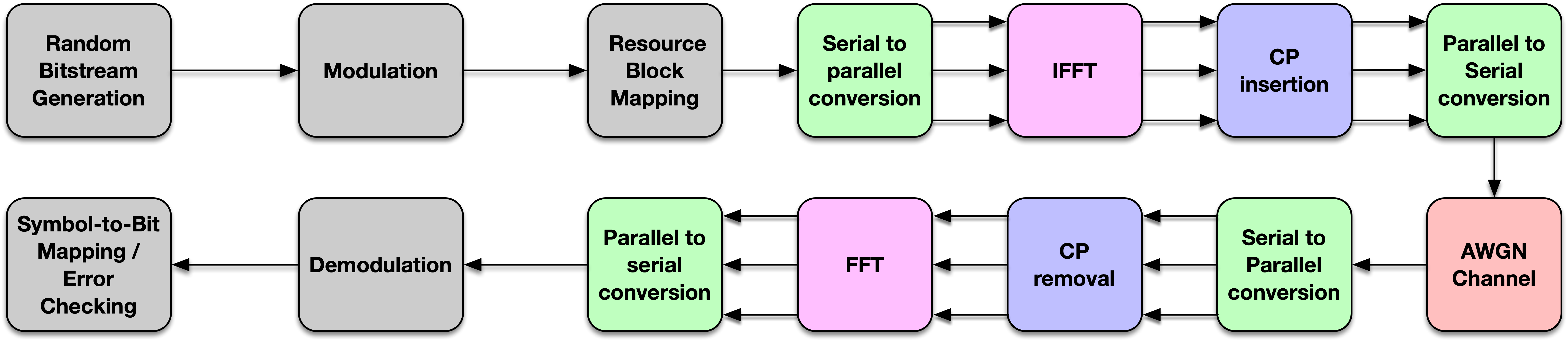}
    \caption{OFDM block diagram of the C++ PHY-fidelity simulator.}
    \label{fig:OFDM}
    \vspace{-5pt}
\end{figure*}

To support scalable training, we implement a high-fidelity OFDM simulator in C++ (Fig.~\ref{fig:OFDM}) that generates link-layer throughput for a single downlink user under configurable PRB allocations, MCS levels, and pathloss conditions. This simulator enables policy learning without hardware dependencies or testbed instability.

For each configuration, a single-user OFDM system is instantiated following 3GPP TS~38.214~\cite{3gpp38214}. We support MCS indices 6–28, spanning QPSK, 16-QAM, and 64-QAM with standard-compliant coding rates. Random information bits are LDPC-encoded, modulated, mapped to allocated PRBs, and converted to the time domain via OFDM with an appropriately sized IFFT and cyclic prefix.

Transmission occurs over a deterministic flat-fading channel with additive white Gaussian noise (–80~dB), isolating pathloss as the dominant impairment. At the receiver, symbols are demodulated, decoded, and compared against transmitted bits to compute throughput as the number of correctly recovered information bits per OFDM symbol duration.

The simulator operates interactively with a Python-based RL agent. At each step, the agent queries the simulator with a PRB allocation, MCS index, and pathloss value and receives the corresponding throughput response. This closed-loop interaction enables efficient exploration of the PRB–throughput–channel relationship across a wide operating range, supporting robust policy learning beyond static offline datasets.

\section{Methodology}
\label{sec:method}

\subsection{Simulation Setup}

We design a simulation environment that captures realistic traffic patterns, SLA constraints, and slice-specific QoS requirements for URLLC, eMBB, and mMTC services. Each training episode consists of $256$ decision steps, during which the agent jointly allocates PRBs to all active slices based on instantaneous demand and service objectives. The episode includes up to 14 UEs: $N_{\text{URLLC}}=3$, $N_{\text{eMBB}}=3$, and $N_{\text{mMTC}}=8$. The agent observes the state of each slice and outputs a single action specifying PRB allocations for URLLC, eMBB, and mMTC. No allocation hierarchy is imposed; instead, the agent learns to balance latency, throughput, and connectivity objectives across slices to maximize cumulative reward.

\subsection{Parameterization of OAI Testbed and Simulator Environments}
\label{subsec:parameterization}

\subsubsection{OAI-Based Testbed Configuration}
\label{config}

The empirical component of MORPH leverages an OAI-based 5G testbed configured with the following parameters:
\begin{itemize}
    \item \textbf{Coverage Area:} $1.5 \times 1.5$ km, with the gNodeB at the center.
    \item \textbf{Carrier Frequency:} NR band n78 ($f_c=3.6$ GHz).
    \item \textbf{Transmission Mode:} TDD with a 6-slot periodicity, comprising 7 DL slots, 2 UL slots, and a mixed slot with 6 DL and 4 UL symbols.
    \item \textbf{Resource Grid:} $106$ PRBs at 30~kHz subcarrier spacing.
\end{itemize}

\subsubsection{Received Power Model in OAI}

In OAI, the reported “pathloss” corresponds to effective received power rather than pure propagation attenuation. Following the OAI slicing model, the link budget incorporates transmit power and antenna gains with the free-space pathloss expression:
\[
P_{\mathrm{rx}} = P_{\mathrm{tx}} + G_{\mathrm{tx}} + G_{\mathrm{rx}} - \big(20\log_{10}(d) + 20\log_{10}(f_c) + 32.44\big),
\]
where $P_{\mathrm{tx}}$ is the gNodeB transmit power, $G_{\mathrm{tx}}$ and $G_{\mathrm{rx}}$ are antenna gains, $d$ is the UE–gNodeB distance (km), and $f_c$ is the carrier frequency (MHz). For NR band n78, the combined gain term is approximately $83.84$~dB, yielding realistic 5G signal levels. Accordingly, the received power is bounded as
\[
P_{\mathrm{rx}} \in [-23.0,\,-7.0]~\mathrm{dB},
\]
% \ba{Should it be dB or dBm?}
% \ts{The FSPL expression we have is in dB as well Tx-and Rx-Gains. Thus we assume PTx in dB as well and PRx becomes dB.}
% \ba{Why are we calling this received power? Shouldn't this be pathloss or channel gain?}
% \ts{Channel gain is gain from the wireless channel, pathloss is the amount of power you loose at a specific distance} \rb{Yes, it's an OAI-ism and Tolunay's explaination is correct for what is in rfsim; which drove a patch}
spanning weak to strong signal regimes. This normalized metric is included in the RL observation space (Section~\ref{subsec:spaces}). We use $P_{\mathrm{rx}}$ as a relative channel-quality input to the policy and normalize it in the observation space; absolute calibration is not required for the comparative training analysis in this work.

\subsubsection{Physical Layer Simulator}

The synthetic training environment employs a high-fidelity OFDM simulator implemented in C++, adhering to 3GPP NR specifications. The simulator models PRB-level allocation, MCS indices 6–28 from TS~38.214~\cite{3gpp_ts38214_mcs_table}, and an AWGN noise floor of $-80$~dB. Throughput is computed from correctly decoded information bits, providing a realistic approximation of achievable link performance without hardware constraints.

\subsubsection{Traffic Models and SLA Configurations}
\label{sec:par:sla}

\begin{itemize}
    \item \textbf{URLLC:} Packet arrivals follow a Bernoulli process with probability $p=0.8$. Packet sizes are sampled as
    \[
    B_{\mathrm{URLLC}} \sim \mathcal{U}(1.5,4.0)\:\mathrm{Mbits},
    \]
    with latency $L_u$ constrained by the SLA $L_u \leq 400$~ms.
    
    \item \textbf{eMBB:} Throughput demand is drawn from
    \[
    T_{\mathrm{req}} \sim \mathcal{U}(5,15)\:\mathrm{Mbps},
    \]
    with step-wise variation sampled from $\mathcal{N}(\mu,0.1\mu)$, where $\mu = T_{req}$. The SLA enforces a minimum throughput of $5$~Mbps. 
    % \fa{define $\mu$}
    
    \item \textbf{mMTC:} Sporadic IoT transmissions are modeled via a connectivity-based service check. Each device requires a minimum throughput of $T_{\mathrm{mMTC}}=3.5$~Mbps, and the number of supported devices is computed as the total allocated mMTC throughput divided by this value. The SLA enforces a minimum 95\% service ratio.
\end{itemize}

\subsection{RL Algorithm Selection and Design Considerations}

We adopt Proximal Policy Optimization (PPO) due to its training stability, sample efficiency, and suitability for constrained action spaces. PPO’s clipped surrogate objective enables smooth policy updates, mitigates instability, and allows multiple updates per batch, making it well suited for MORPH’s hybrid sim–real training setup and slice-aware PRB allocation task.

\subsection{Action and Observation Space Specification}
\label{subsec:spaces}

\subsubsection{Observation Space}

The observation space is a continuous vector encoding slice identity, normalized traffic demand, and normalized received power for each UE. For $N_{\mathrm{UE}}$ users, the observation dimension is $3N_{\mathrm{UE}}$:
\[
\mathbf{o}_i = [\text{s\_id}_i,\;\hat{r}_i,\;\hat{P}_{\mathrm{rx},i}],
\]
where $\text{s\_id}_i \in \{0,1,2\}$ denotes the slice type corresponding to URLLC, eMBB, or mMTC, respectively. The normalized traffic demand $\hat{r}_i$ is defined as
\[
\hat{r}_i = \frac{r_i - r_{\min}^{(s)}}{r_{\max}^{(s)} - r_{\min}^{(s)}}, 
\quad s \in \{\text{URLLC},\text{eMBB}\},
\]
where $r_i$ denotes the instantaneous traffic demand (or requested data rate) of UE~$i$, and $r_{\min}^{(s)}$ and $r_{\max}^{(s)}$ are slice-specific demand bounds determined by the corresponding traffic models.

\subsubsection{Action Space}

The action space consists of PRB allocation tuples
\[
\mathbf{a} = (\text{PRB}_{\mathrm{URLLC}},\text{PRB}_{\mathrm{eMBB}},\text{PRB}_{\mathrm{mMTC}}),
\]
subject to the resource constraint
\[
\text{PRB}_{\mathrm{URLLC}} + \text{PRB}_{\mathrm{eMBB}} + \text{PRB}_{\mathrm{mMTC}} = 106.
\]

\section{Data Quality and Model Validation}
\label{sec:data-quality}

%% Basir Start
\subsubsection{Empirical MCS Profiling in OAI}
To ground throughput modeling in standards-compliant link adaptation, we log the downlink MCS indices selected by OAI over extended runs across the received-power range. For each $P_{\mathrm{rx}}$ point, we compute an empirical histogram $P(\text{mcs}\mid P_{\mathrm{rx}})$, which captures the stochastic behavior induced by block error rate (BLER)-targeted adaptation and implementation effects in the OAI stack. We use an AWGN channel in RF-simulator mode as a controlled baseline to isolate link-adaptation behavior from multipath; extending profiling to frequency-selective fading is an important direction for future work, but is outside the scope of the current experimental campaign.

%% Basir End

\subsubsection{Theoretical Throughput Calculation}

The theoretical throughput, $T_{\text{theoretical}}$, is computed by combining physical-layer parameters with empirical MCS distributions extracted from OAI logs:
\[
T_{\text{theoretical}}(P_{\mathrm{rx}}, N_{\text{PRBs}}) 
= \sum_{mcs=6}^{28} P(mcs \mid P_{\mathrm{rx}})\, T_{mcs}(N_{\text{PRBs}}),
\]
where $P(mcs \mid P_{\mathrm{rx}})$ denotes the empirically observed probability of selecting MCS index $mcs$ at received power $P_{\mathrm{rx}}$, and $T_{mcs}(N_{\text{PRBs}})$ represents the achievable throughput under that MCS and PRB allocation.

For a given MCS, the throughput contribution is computed as
\[
\begin{split}
T_{mcs}(N_{\text{PRBs}}) 
&= \frac{N_{\text{PRBs}}}{106}
\times N_{\text{sc}} \times N_{\text{symb}} \times N_{\text{slots}} \quad \times Q_m \times R_m \times \\
&N_{\text{layers}}
\times \eta_{\text{DL}} \times (1 - \eta_{\text{OH}}) \times (1 - \text{BLER}),
\end{split}
\]
where $N_{\text{sc}}=12$ is the number of subcarriers per PRB, $N_{\text{symb}}=14$ is the number of OFDM symbols per slot, and $N_{\text{slots}}=2000$ denotes the number of downlink slots per second under the configured numerology. The modulation order $Q_m \in \{2,4,6\}$ and coding rate $R_m$ correspond to the selected MCS index according to 3GPP TS~38.214. $N_{\text{layers}}$ denotes the number of transmission layers, $\eta_{\text{DL}}=\frac{104}{140}$ accounts for the downlink transmission duty cycle based on the configuration in Section~\ref{config}, $\eta_{\text{OH}}=0.14$ captures control and protocol overhead, and BLER represents the block error rate associated with the selected MCS.

\subsubsection{Practical Throughput Calculation}
Practical throughput uses application-level measurements from OAI-based iPerf tests:
\[
T_{\text{practical}}(P_{\mathrm{rx}}, N_{\text{PRBs}}) = T_{\text{iperf}}(P_{\mathrm{rx}}) \times \frac{N_{\text{PRBs}}}{106},
\]
where $T_{\text{iperf}}(P_{\mathrm{rx}})$ captures protocol overhead, HARQ delays, and realistic scheduler behavior.

\subsubsection{Data Source Validation}
\begin{figure}[h]
    \centering
    \includegraphics[width=0.85\linewidth]{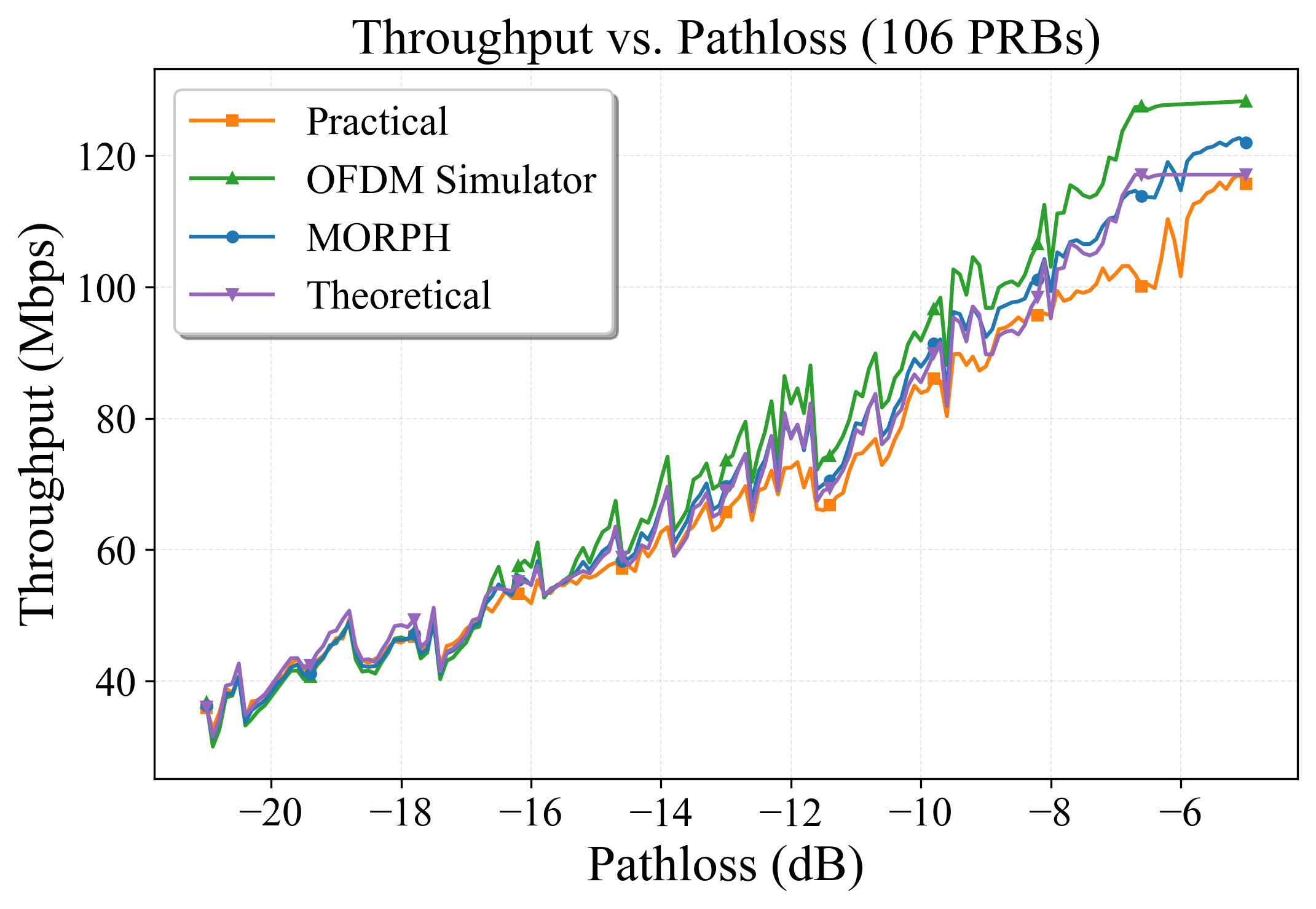}
    \caption{Throughput vs Received Power: Comparison of theoretical and practical methods for 106 PRBs. %\fa{let's say MORPH instead of hybrid in the legend}
    }
    \label{fig:throughput-comparison}
\end{figure}

To validate data quality, we analyze consistency between theoretical and practical throughput across the full received power range. Fig.~\ref{fig:throughput-comparison} demonstrates a strong correlation, confirming that both models exhibit the expected throughput degradation trend as received power decreases (equivalently, as effective pathloss increases), with deviations attributable to protocol overheads and BLER-targeted link adaptation.

\subsubsection{MCS Distribution Analysis}
\begin{figure}[h]
    \centering
    \includegraphics[width=0.8\linewidth]{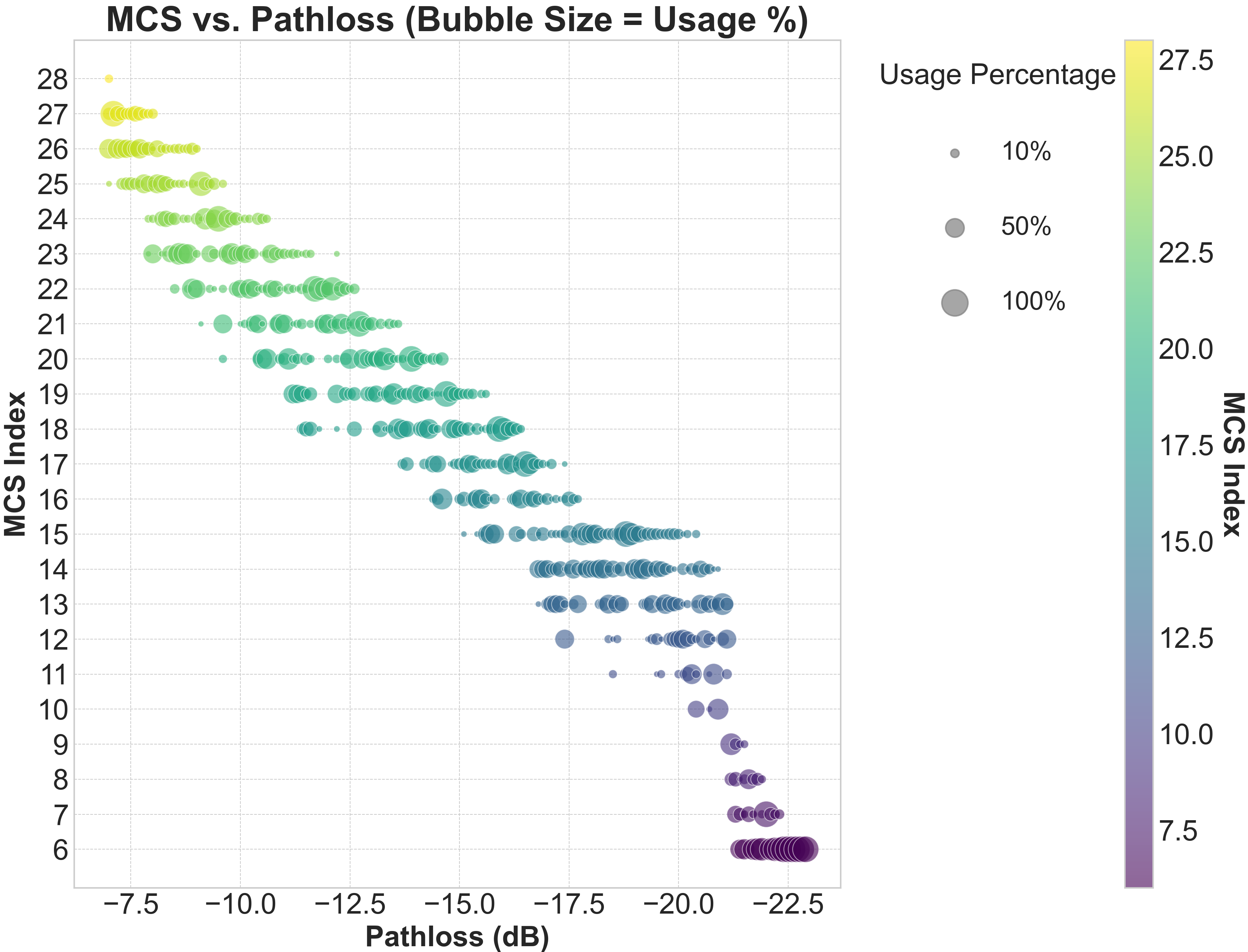}
    \caption{MCS vs Received Power: Bubble chart shows adaptive modulation and coding in OAI across the channel quality range. %\fa{numbers are not readable. can you enlarge and bold the font in the color code and axis}
    }
    \label{fig:mcs-vs-pathloss}
\end{figure}

Fig.~\ref{fig:mcs-vs-pathloss} shows the adaptive modulation pattern extracted from OAI logs. Bubble sizes indicate the frequency of MCS usage, forming the empirical foundation for the weighted-average approach in theoretical calculations.

\subsubsection{Conclusion}
The alignment between theoretical and practical calculations, combined with empirical MCS distributions, establishes the reliability of both data sources for hybrid RL training.

\section{Evaluation}
\label{sec:evaluation}

To contextualize the evaluation results, we summarize the defining characteristics of the three training environments in Table~\ref{tab:env_comparison}. The table highlights how agents differ in throughput measurement, state exploration, execution realism, and scalability across the OAI-based, OFDM-fidelity, and hybrid (MORPH) setups.

We evaluate each agent based on its ability to allocate PRBs to maximize application-layer throughput under varying channel quality (parameterized by $P_{\mathrm{rx}}$, i.e., the RF-simulator pathloss knob). All agents are evaluated online in the OAI RF-simulator harness using \texttt{iPerf}-measured throughput; the throughput estimators are used only during offline training to avoid exhaustive OAI sweeps over the full PRB--channel state space.

\noindent\textbf{Baseline 1 (Practical Agent)} is trained using application-layer throughput measurements collected from the OAI platform via \texttt{iPerf}. Throughput obtained under full PRB allocation (106 PRBs) is used as a reference, and throughput for other PRB configurations is computed proportionally. This approach is necessary because exhaustive data collection over the full PRB state space is computationally infeasible.

\noindent\textbf{Baseline 2 (Simulated Agent)} relies on throughput generated by the OFDM simulator. For each pathloss value, average throughput is computed as a weighted sum over MCS levels, where each MCS contribution is weighted by its empirical occurrence probability (Fig.~\ref{fig:mcs-vs-pathloss}).

\noindent\textbf{The MORPH Agent (Hybrid)} employs a \textit{Hybrid Reward Ensemble} to mitigate the bias-variance trade-off inherent in Sim-to-Real transfer. We define the reward signal $R(s,a)$ as a weighted ensemble of the empirical and synthetic estimators:
\begin{equation}
    R_{hybrid}(s,a) = \lambda \cdot \hat{T}_{OAI}(s,a) + (1-\lambda) \cdot \hat{T}_{PHY}(s,a)
\end{equation}
where $\hat{T}_{PHY}$ is the throughput derived from the idealized 3GPP simulator, offering a low-variance but biased learning signal (it omits protocol-stack overheads), and $\hat{T}_{OAI}$ is the throughput derived from empirical OAI measurements, which is higher-variance but captures stack-level effects and implementation constraints.
 This formulation acts as a regularizer: the simulator encourages exploration, while the OAI data penalizes allocations that are theoretically optimal but practically unachievable due to stack overheads. For this work, we set $\lambda = 0.5$.

\textbf{Evaluation execution} is performed using the OAI RF-simulator testbed under the serialized UE harness described in Section~III-A. For fair comparison, all agents are evaluated under identical scenario traces (traffic demands and received-power evolution). The interpolated throughput tables are used \emph{only} during off-testbed training to avoid exhaustive OAI sweeps over the full PRB--channel state space.

\begin{figure*}[htbp]
    \centering
    \includegraphics[width=0.9\linewidth]{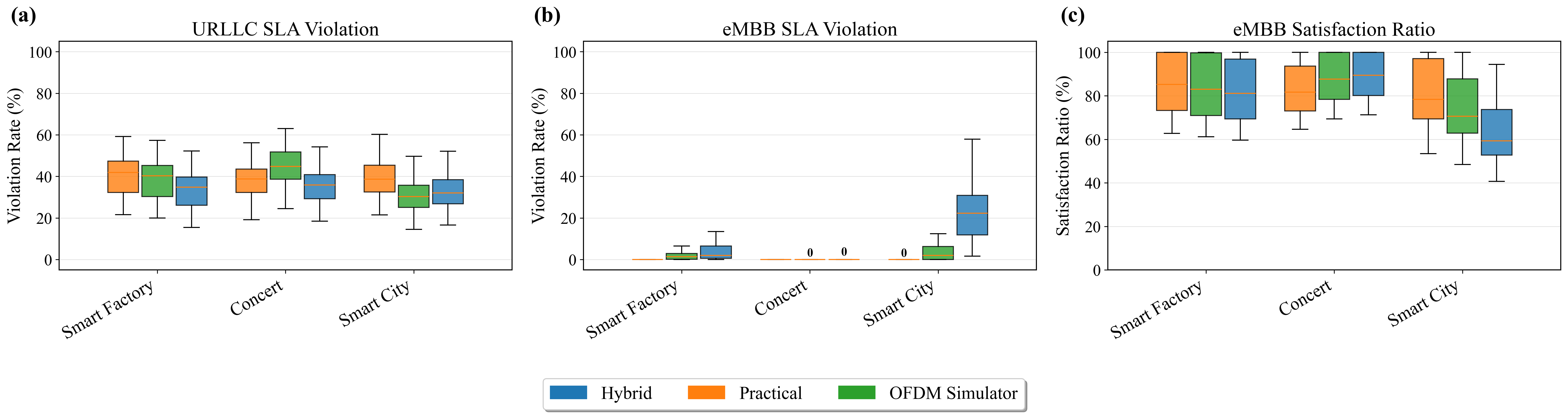}
    \caption{SLA violation and satisfaction metrics across URLLC and eMBB services for different agent types (Practical, Simulated, Hybrid) under Smart Factory, Concert, and Smart City scenarios.}
    \label{fig:sla-violation}
\end{figure*}

\begin{figure*}[htbp]
    \centering
    \includegraphics[width=0.9\linewidth]{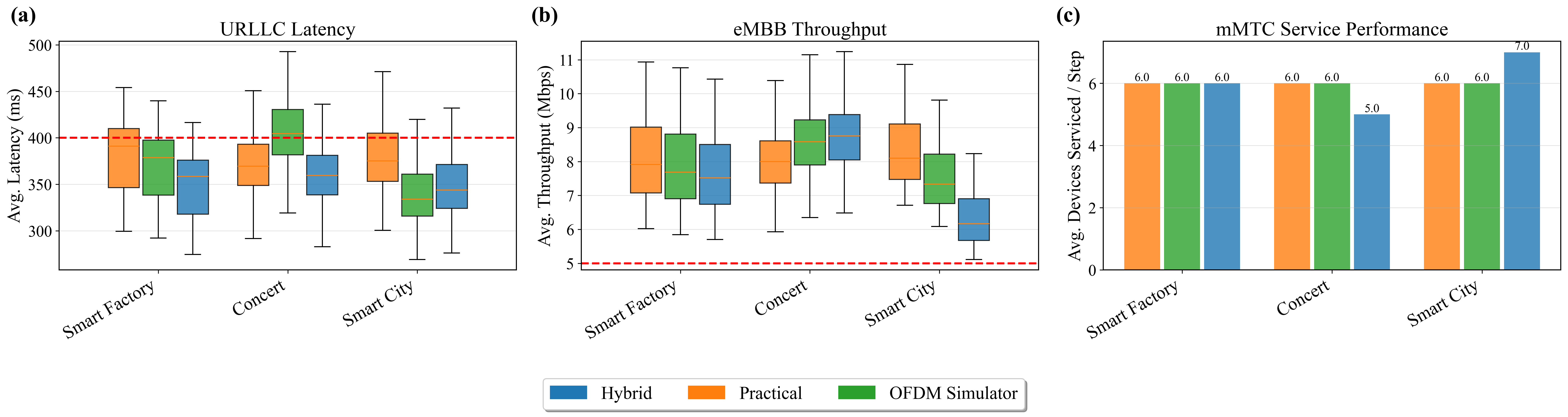}
    \caption{Average latency, throughput, and mMTC service performance across agents and scenarios. Red dashed lines indicate SLA thresholds.}
    \label{fig:latency-throughput}
\end{figure*}

\begin{figure*}[htbp]
    \centering
    \includegraphics[width=0.9\linewidth]{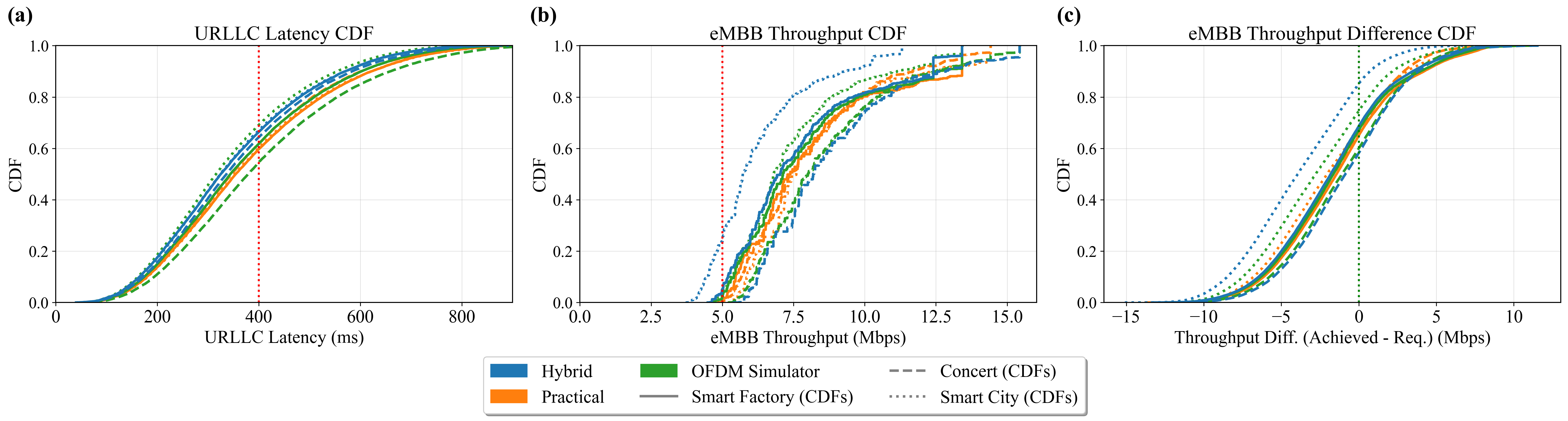}
    \caption{CDF analysis of URLLC latency and eMBB throughput for Practical, Simulated, and Hybrid agents. Vertical lines denote key thresholds.}
    \label{fig:cdf-analysis}
\end{figure*}

\section{Results and Discussion}

The performance of the hybrid agent and the two baselines is evaluated across three representative scenarios, with results shown in Figs.~\ref{fig:sla-violation}–\ref{fig:cdf-analysis}. These results demonstrate MORPH’s ability to learn scenario-adaptive PRB allocation policies under heterogeneous QoS demands.

\begin{itemize}[leftmargin=*]
    \item \textbf{Smart Factory (URLLC-Centric):} The hybrid agent achieves the lowest average URLLC latency, consistently remaining below the 400~ms SLA threshold, and exhibits significantly fewer SLA violations than the baselines (Figs.~\ref{fig:latency-throughput}(a), \ref{fig:sla-violation}(a)). This improved URLLC performance comes at the cost of reduced eMBB throughput and satisfaction (Figs.~\ref{fig:latency-throughput}(b), \ref{fig:sla-violation}(c)), reflecting the scenario’s URLLC-weighted reward configuration ($w_{\mathrm{URLLC}}=0.4$).

    \item \textbf{Stadium (eMBB-Centric):} Under eMBB prioritization, the hybrid agent attains the highest average eMBB throughput, near-zero SLA violation rates, and the highest user satisfaction (Figs.~\ref{fig:latency-throughput}(b), \ref{fig:sla-violation}(b), \ref{fig:sla-violation}(c)). This is achieved by reallocating resources away from mMTC (Fig.~\ref{fig:latency-throughput}(c)) while maintaining URLLC performance, demonstrating effective multi-objective trade-offs.

    \item \textbf{Smart City (mMTC-Centric):} In the mMTC-focused scenario, the hybrid agent supports the largest number of connected IoT devices (Fig.~\ref{fig:latency-throughput}(c)), prioritizing connectivity as the primary objective. This is accomplished by degrading eMBB performance while minimally affecting URLLC, consistent with the reward weighting ($w_{\mathrm{mMTC}}=0.4$).
\end{itemize}

\subsection{Limitations and Implications}
\label{sec:limitations}
Our OAI execution harness uses serialized UE activation to avoid CPU-induced distortions in software baseband processing. This design yields stable and repeatable application-layer throughput measurements but omits simultaneous multi-UE contention and interference effects in the testbed execution path. Similarly, AWGN is used as a controlled channel model for MCS profiling in RF-simulator mode, isolating link-adaptation behavior from multipath. Consequently, MORPH should be interpreted as a methodology for improving the fidelity and coverage of the throughput learning signal under controlled testbed constraints, rather than as a complete evaluation of multi-user scheduling under fully realistic propagation. These limitations do not invalidate the main contribution—measurement-grounded throughput modeling and hybrid training-signal construction—but they do bound the direct generalization claims.

\section{Conclusion}

This paper introduced MORPH, a multi-environment orchestrated RL framework that bridges the simulation-to-reality gap in O-RAN radio resource management. By combining empirical measurements from an OAI testbed with interactive exploration in a high-fidelity physical-layer simulator, MORPH enables robust and generalizable PRB allocation policies. Across the considered scenarios, MORPH improves robustness of slice-wise performance relative to single-source training baselines, with the clearest gains appearing in regimes where protocol-stack effects and PHY-only modeling diverge. While MORPH does not dominate every metric under every scenario configuration, it provides a more reliable throughput learning signal for offline policy optimization under controlled OAI execution. Extending the evaluation beyond serialized execution and beyond AWGN to incorporate stronger multi-user contention and frequency-selective fading remains an important direction for future work.

\FloatBarrier        
\nocite{*}
\bibliographystyle{IEEEtran}
\bibliography{References}
\end{document}